\pdfoutput=1
\documentclass[journal=jpclcd, layout=twocolumn, manuscript=letter]{achemso}
\usepackage[T1]{fontenc}
\usepackage{amsmath}
\newcommand*{\angstrom}{\textup{\AA}}
\usepackage{booktabs}
\usepackage{graphicx}
\usepackage{color}
\usepackage{natbib}
\usepackage{natmove}
\newcommand*{\onlinecite}[1]{\hspace{-1 ex} \nocite{#1}\citenum{#1}}

\title{Structural Complexity and Phonon Physics in 2D Arsenenes}

\author{Jes\'{u}s Carrete}
\affiliation{Institute of Materials Chemistry, TU Wien, A-1060 Vienna, Austria}
\email{jesus.carrete.montana@tuwien.ac.at}

\author{Luis J. Gallego}
\affiliation{Departamento de F\'isica de la Materia Condensada, Facultad de F\'isica, Universidad de Santiago de Compostela, E-15782 Santiago de Compostela, Spain}

\author{Natalio Mingo}
\affiliation{Universit\'e Grenoble Alpes, F-38000 Grenoble, France\\
  CEA, LITEN, 17 rue des Martyrs, F-38054 Grenoble, France}

\begin{tocentry}
  \includegraphics{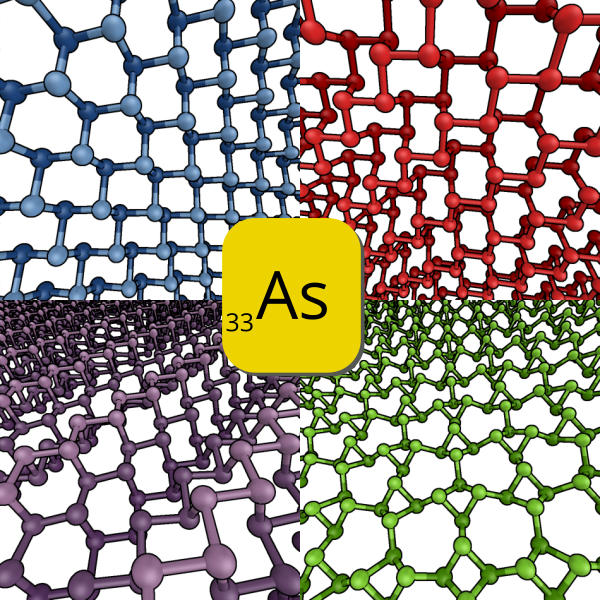}
\end{tocentry}

\begin{document}

\begin{abstract}
  In the quest for stable 2D arsenic phases, four different structures have been recently claimed to be stable. We show that, due to phonon contributions, the relative stability of those structures differs from previous reports and depends crucially on temperature. We also show that one of those four phases is in fact mechanically unstable. Furthermore, our results challenge the common assumption of an inverse correlation between structural complexity and thermal conductivity. Instead, a richer picture emerges from our results, showing how harmonic interactions, anharmonicity and symmetries all play a role in modulating thermal conduction in arsenenes. More generally, our conclusions highlight how vibrational properties are an essential element to be carefully taken into account in theoretical searches for new 2D materials.
\end{abstract}

After the first experimental isolation and electronic characterization of graphene \cite{novoselov_electric_2004}, growing interest has been devoted to the properties of two-dimensional (2D) nanostructures. Theoretical predictions have often preceded synthesis and characterization due to the difficulties of nanoscale experimental measurements. Due to their peculiar physical and chemical properties, 2D nanostructures have a huge diversity of potential applications in nanoelectronics, spintronics, optoelectronics, nanomedicine, hydrogen storage and so forth \cite{novoselov_two-dimensional_2005, geim_rise_2007, castro_neto_electronic_2009, chen_graphene-based_2010, das_sarma_electronic_2011, singh_graphene_2011, sun_graphene_2011, huang_overview_2012, wang_electronics_2012, geim_van_2013, butler_progress_2013, xu_graphene-like_2013, gupta_recent_2015}. At the same time, the number of already synthetized or predicted 2D materials is increasing continually.

Technological applications of group-IV 2D phases such as graphene \cite{novoselov_electric_2004}, silicene and germanene \cite{yan_electron-phonon_2013} are constrained by their zero band gap, specially in optoelectronics. Although the gap can be opened by doping \cite{balog_bandgap_2010, quhe_tunable_2012, ye_tunable_2014} or via an external electric field \cite{mak_observation_2009, ni_tunable_2012}, this has spurred recent interest in other 2D materials. Those derived from group-V elements show particular potential because of the quasi-layered character of their bulk forms. Single-layer black phosphorene was investigated early on \cite{liu_phosphorene:_2014}, and found to have a wide $\sim 2\;\mathrm{eV}$ band gap vs. the $\sim 0.34\;\mathrm{eV}$ gap of the bulk \cite{li_black_2014}. Black phosphorene is puckered, rather than flat (as graphene) or buckled (as silicene). However, buckled (blue) phosphorene was also theoretically predicted to be stable \cite{zhu_semiconducting_2014}. Subsequent calculations \cite{zhang_atomically_2015} showed that 2D honeycomb structures of As and Sb (both semimetallic in bulk form) could be of interest for optoelectronics, as both are indirect semiconductors with indirect-direct band-gap transitions under mild biaxial strain.  Almost at the same time, Kamal and Ezawa \cite{kamal_arsenene:_2015} theoretically predicted the stability of buckled and puckered arsenene structures with strain-tunable indirect band gaps, and Zhang et al. \cite{zhang_manifestation_2015} and Han et al. \cite{han_negative_2015} showed that multilayer orthorhombic arsenenes behave as intrinsic direct bandgap semiconductors with gap values of $\sim 1\;\mathrm{eV}$, high carrier mobilities and negative Poisson's ratios. Recently, extensive work by Ciraci and coworkers on the mechanical and electronic properties of nitrogene \cite{ozcelik_prediction_2015}, arsenene \cite{kecik_stability_2016}, antimonene \cite{akturk_single-layer_2015} and bismuthene \cite{akturk_single_2016} has complemented earlier studies on phosphorene \cite{liu_phosphorene:_2014, li_black_2014, zhu_semiconducting_2014} to provide a complete picture of these properties for group-V 2D materials.

A fundamental element in all these theoretical inquiries has been the stability of the predicted monolayers, both with respect to small perturbations (mechanical stability) and to possible competing 2D phases. Mechanical stability has been investigated by calculating the phonon spectra from first principles, and sometimes also by finite-temperature molecular dynamics\cite{kecik_stability_2016, ersan_stable_2016}. Regarding thermodynamic stability, it is typically evaluated from accurate first-principle estimates of the energy per atom in each structure. This approximation neglects the vibrational contribution to the free energy, important even at zero temperature. Furthermore, conclusions are often compromised by a failure to enforce the fundamental rotational invariance of mechanics upon the phonon dispersions, which is critical in 2D materials \cite{carrete_physically_2016}.

\begin{figure}
  \begin{center}
    \includegraphics[width=\linewidth]{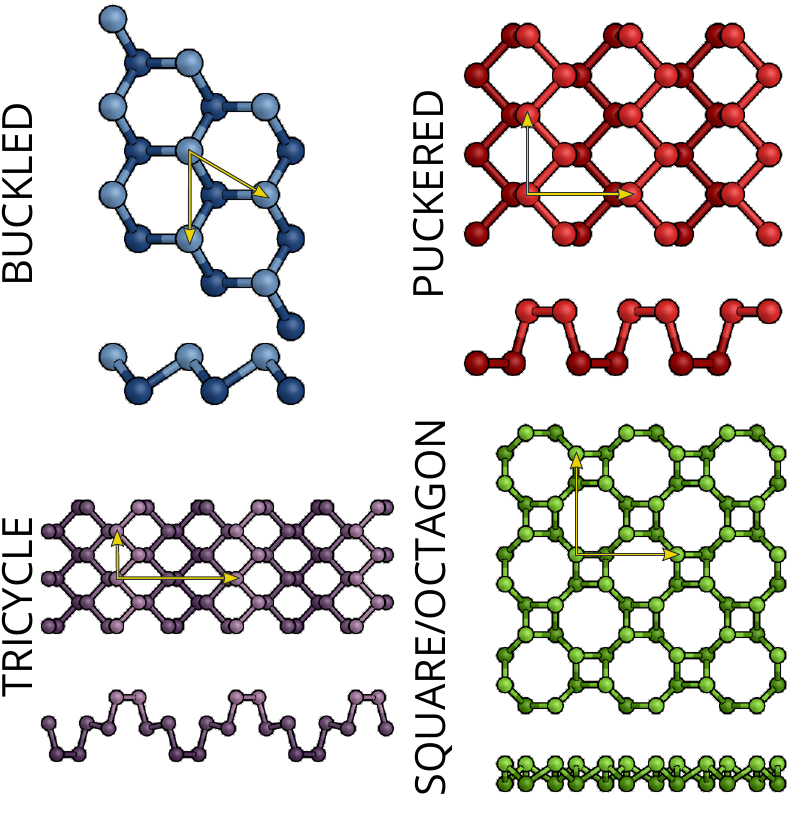}
  \end{center}
  \caption{Top and side views of a $3\times 3$ supercell of each of the arsenene structures considered in this paper. The superimposed arrows represent the unit cell vectors.}
  \label{fig:kugel-stab}
\end{figure}

For arsenene, both a buckled honeycomb and a puckered structure were originally proposed \cite{zhang_atomically_2015, kamal_arsenene:_2015, kecik_stability_2016}. However, very recently two new candidate structures have been put forward: a tricycle structure \cite{ma_two-dimensional_2016}, and a structure consisting of buckled square and octagon rings \cite{ersan_stable_2016}. Both have been predicted to be direct-band-gap semiconductors with appreciable gaps. Multilayer arsenene nanoribbons have already been synthesized on an InAs substrate using the plasma-assisted process, and the band gap has been estimated to be about $2.3\;\mathrm{eV}$ \cite{tsai_direct_2016}. All four structures are depicted in Fig. \ref{fig:kugel-stab}.

The overall theoretical description of arsenene has recently been supplemented by two studies on its thermal conductivity \cite{zeraati_highly_2016, zheng_comparative_2016}, which is also determined by the physics of phonons in the material. Beyond its singular theoretical appeal, thermal transport in 2D materials is important with a view to practical applications such as heat dissipation in nanoelectronics. The first of those published studies \cite{zeraati_highly_2016} is focused on the puckered structure of arsenene, finding a very anisotropic room-temperature thermal conductivity of $30.4$ and $7.8\;\mathrm{W / \left(m\;K\right)}$ along the zigzag and armchair directions, respectively. The second  \cite{zheng_comparative_2016} contains a comparison between the thermal conductivities of the (isotropic) buckled and (anisotropic) puckered structures. Both were performed using standard ab-initio methods and the ShengBTE solver of the Boltzmann transport equation for phonons \cite{cpc_2014}. However, inspection of their computed phonon spectra reveals possible violations of rotational symmetry in their force constants, which can strongly influence the predicted thermal conductivities (to the point of reversing their anisotropy) as shown for  borophene \cite{carrete_physically_2016}. Specifically, the telltale sign of such violation is the lack of a quadratic ZA branch  with zero group velocity close to the $\Gamma$ point.

Here we perform a first-principles study of the equilibrium structures and lattice dynamics of the four possible phases of arsenene represented in Fig. \ref{fig:kugel-stab}: buckled, puckered, tricycle and square/octagon (s/o). We find that the tricycle phase is mechanically unstable and discard it for further study. More importantly, we show that inclusion of phonons in the energetic picture turn the puckered and s/o structures into the most stable ones at low and room temperatures, respectively. This contradicts previous analyses based on ground-state energies alone, which predict the buckled structure as the most stable one. We also study thermal transport in all four structures, and show a lack of a simple correlation between unit-cell complexity and thermal conductivity.

We start by performing a local relaxation of each structure using the density functional theory (DFT) package VASP \cite{vasp_general_1, vasp_general_2, vasp_general_3, vasp_general_4} with projector-augmented-wave datasets \cite{vasp_paw_1, vasp_paw_2} and the Perdew-Burke-Ernzerhof approximation to exchange and correlation \cite{vasp_pbe_1, vasp_pbe_2}. We achieve fully converged results employing a $40\times 40$ $\Gamma$-centered regular grid in reciprocal space, and a cutoff energy of $261\;\mathrm{eV}$. Our unit cells span $25\;\angstrom$ in the through-plane directions of the layers to avoid spurious interactions due to the periodic boundary conditions. The main geometric features obtained from the relaxation are summarized in Table \ref{tbl:geometry} along with their cohesive energies per atom, defined as the difference between the energy of an isolated As atom and the average energy per atom in the structure.

\begin{table*}
  \caption{Main geometric features and cohesive energies of the four arsenene structures after relaxation.}
  \label{tbl:geometry}
  \begin{center}
  \begin{tabular}{ccccccc}\toprule
    structure & \vtop{\hbox{\strut space}\hbox{\strut group}} & \vtop{\hbox{\strut lattice}\hbox{\strut parameters $(\angstrom )$}} & \vtop{\hbox{\strut atoms /}\hbox{\strut unit cell}} & \vtop{\hbox{\strut bond}\hbox{lengths  $(\angstrom )$}} & \vtop{\hbox{\strut frozen-nuclei}\hbox{\strut cohesive}\hbox{\strut energy}\hbox{\strut ($\mathrm{eV / atom}$)}} &
\vtop{\hbox{\strut zero-point}\hbox{\strut harmonic}\hbox{\strut energy}\hbox{\strut ($\mathrm{eV / atom}$)}}\\\midrule
    buckled & $\mathrm{P\bar{3}m1}$  & $3.607$ & $2$ & $2.508$ & $2.963$ & $2.763$\\
    puckered &$\mathrm{Pmna}$ & $3.686$, $4.754$ & $4$ & $2.492$, $2.514$ & $2.926$ & $2.711$\\
    tricycle & $\mathrm{Pbcm}$ & $3.646$, $9.543$ & $8$ & $2.500$, $2.512$ & $2.947$ & unstable\\
    s/o & $\mathrm{P4/nbm}$ & $7.126$ & $8$ & $2.504$, $2.525$ & $2.862$ & $2.699$\\\bottomrule
  \end{tabular}
  \end{center}
\end{table*}

Our geometric results for the buckled, puckered and s/o structures agree quite well with the results reported in Ref. \onlinecite{ersan_stable_2016}, although our cohesive energies are slightly lower. Likewise, the geometry we obtain for the tricycle structure is in fair agreement with that reported in Ref. \onlinecite{ma_two-dimensional_2016}, and its energetics place it between the buckled and puckered structures.

\begin{figure*}
  \begin{center}
    \noindent\includegraphics[width=\linewidth]{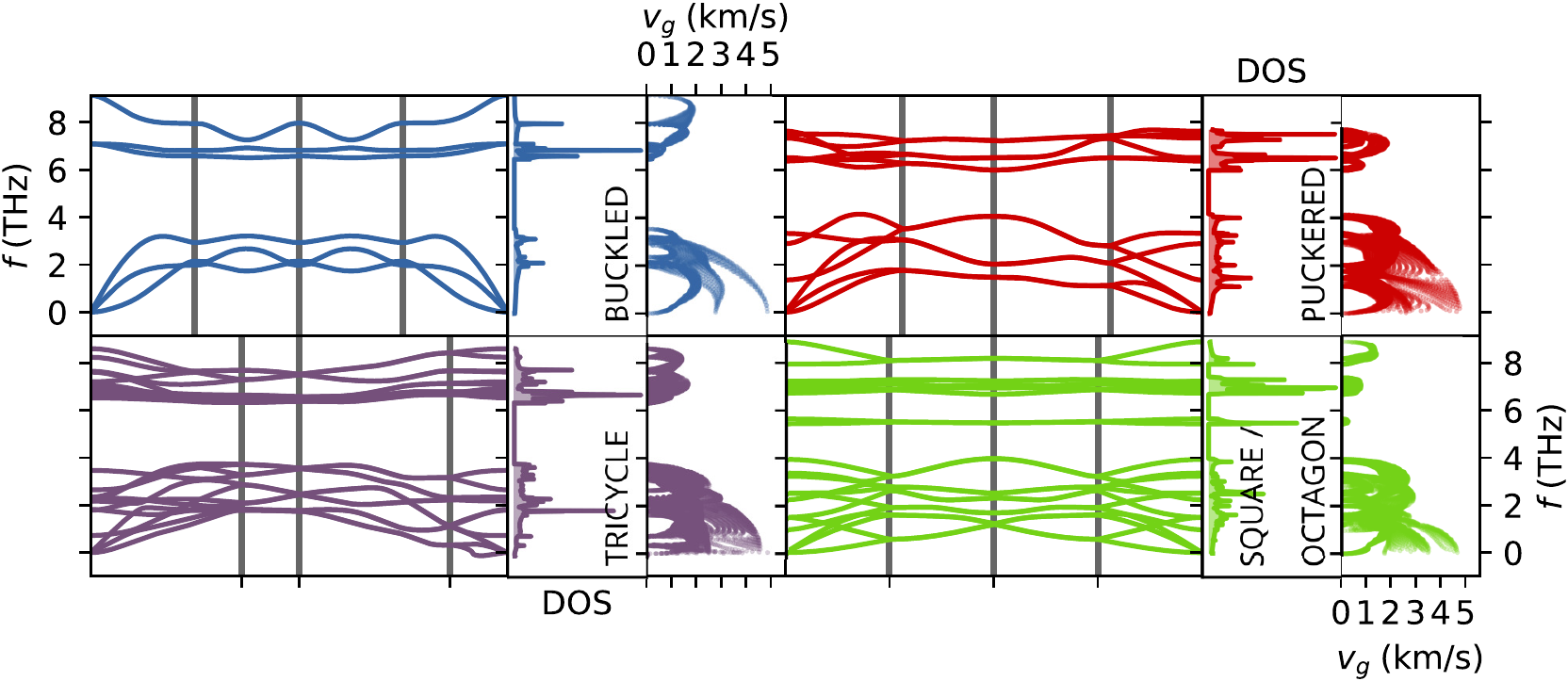}
  \end{center}
  \caption{Phonon dispersions, normalized vibrational densities of states, and group velocities for the four phases of arsenene studied in this work. In each case the bands are plotted along the path $\Gamma\rightarrow \left(1/2, 0\right) \rightarrow \left(1/2, 1/2\right) \rightarrow \left(0, 1/2\right) \rightarrow \Gamma$ in the respective reduced reciprocal coordinates. Imaginary frequencies are represented as negative. Group velocities are calculated on a $100\times 100$ grid.}
  \label{fig:phonons}
\end{figure*}

Our next step is to calculate the phonon dispersions for each structure using a real-space method. We use Phonopy \cite{phonopy} to create a minimal set of displaced supercell configurations and VASP to compute the forces induced by those displacements. We choose supercell sizes of $9\times 9$, $7\times 7$, $5\times 5$ and $5\times 5$ for the buckled, puckered, tricycle and s/o structures respectively. After obtaining a set of ``raw'' DFT-based force constants, we correct them as in Ref. \onlinecite{carrete_physically_2016} to strictly enforce the translation and rotation symmetries of free space. The resulting phonon dispersions, vibrational densities of states, and phonon group velocities are shown in Fig. \ref{fig:phonons}.

All four spectra show an energy gap between a group of high-energy optical modes and the remaining (acoustic plus optical) phonon branches, a general feature of 2D arsenenes according to the literature \cite{zhang_atomically_2015, kamal_arsenene:_2015, kecik_stability_2016, ersan_stable_2016, zeraati_highly_2016, zheng_comparative_2016} with the only exception of Ref. \onlinecite{ma_two-dimensional_2016}, which introduces the tricycle structure. Our prediction for the spectrum of this structure also diverges from Ref. \onlinecite{ma_two-dimensional_2016} in that we predict an imaginary branch extending more than halfway from $\Gamma$ to $\left(0, 1/2\right)$. This feature is resilient to changes in supercell sizes and other parameters. This is the fingerprint of mechanical instability, and therefore the tricycle structure is not included in our remaining calculations.

Reference \onlinecite{ma_two-dimensional_2016} shares another unfortunate feature with most published spectra: mild \cite{zeraati_highly_2016} or serious \cite{zhang_manifestation_2015, zhang_atomically_2015, kamal_arsenene:_2015, kecik_stability_2016, zheng_comparative_2016} violations of the fundamental isotropy of free space reflected in a non-zero group velocity of the convex phonon branch close to $\Gamma$. Only the authors of Ref. \onlinecite{ersan_stable_2016} seem to have enforced full rotational symmetry. As noted in previous studies \cite{carrete_physically_2016, smith_temperature_2017}, this can make the difference between a fully stable spectrum and one containing artifactual imaginary frequencies. This could also be the reason for the prediction of buckled arsenene as unstable in Ref. \onlinecite{kamal_arsenene:_2015}.

Equipped with these phonon spectra, we improve our estimate of the total energy of each structure at $0\;\mathrm{K}$ beyond the frozen-nuclei approximation by including the ground-state energy of each vibrational mode:

\begin{equation}
  E = E_{\mathrm{frozen}} + \frac{h}{2}\sum\limits_{\lambda}f_{\lambda}.
  \label{eqn:0K}
\end{equation}

\noindent Here, $f_{\lambda}$ is the frequency of the vibrational mode in branch $\alpha$ with wave vector $\boldsymbol{q}$, collectively labeled by an index $\lambda$. $\sum\limits_{\lambda}$ denotes both sum over $\alpha$ and an average over the Brillouin zone. This correction (see values in Table \ref{tbl:geometry}) changes the ordering of the cohesive energies of the three stable phases: $0.200\;\mathrm{eV / atom}$ (buckled), $0.216\;\mathrm{eV / atom}$ (puckered) and $0.163\;\mathrm{eV / atom}$ (s/o). Therefore the most stable phase at vanishing temperature is the puckered structure, and not the buckled one.

\begin{figure}
  \begin{center}
    \includegraphics[width=\linewidth]{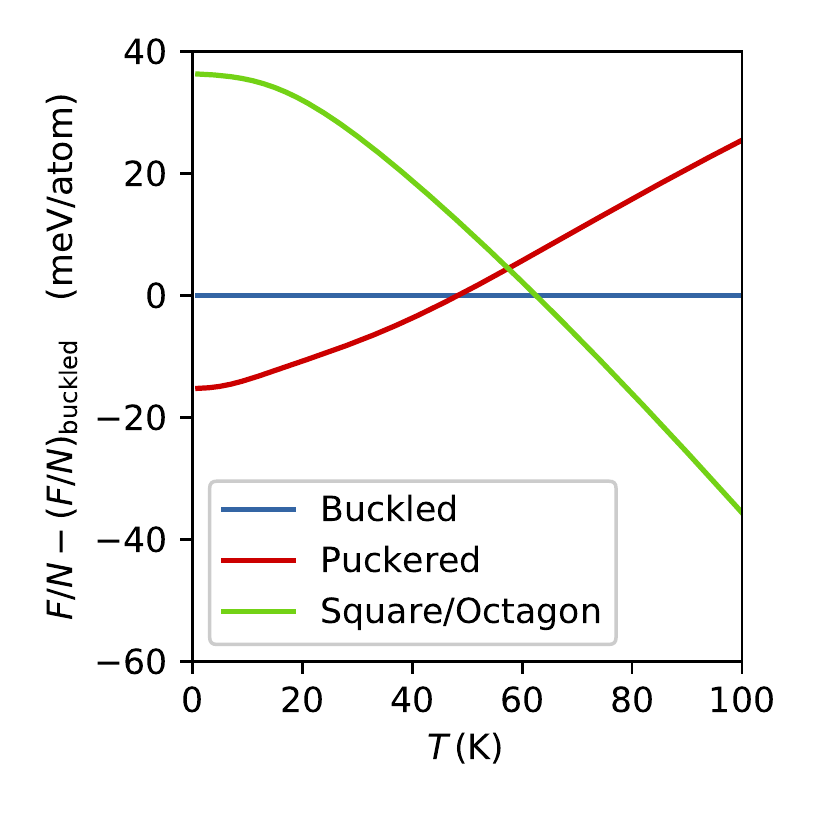}
  \end{center}
  \caption{Relative harmonic free energies of the three stable phases of arsenene studied in this work, as a function of temperature. The free energy of the buckled phase at each temperature is chosen as the reference.}
  \label{fig:free_energy}
\end{figure}

We complete the study of the stability of 2D arsenenes by computing the full harmonic free energy, obtaining by adding the temperature-dependent energy and entropy terms to Eq. (\ref{eqn:0K}):

\begin{align}
  F =& E_{\mathrm{frozen}} + E_{\mathrm{vib}}\left(T\right) - TS_{\mathrm{vib}}\left(T\right) =\nonumber\\
  =&E_{\mathrm{frozen}} - \frac{h}{2}\sum\limits_{\lambda}f_{\lambda} - k_B T\sum\limits_{\lambda} \log n_{\lambda}\left(T\right),
  \label{eqn:harmonic}
\end{align}

\noindent where $n_\lambda\left(T\right) = \left[\exp\left(\frac{h f_{\lambda}}{k_B T}\right) - 1\right]^{-1}$ is the Bose-Einstein occupancy factor of mode $\lambda$ at temperature $T$. The results are plotted in Fig. \ref{fig:free_energy}. The crossover around $60\;\mathrm{K}$ suggests that the square/octagon phase \cite{ersan_stable_2016} is the thermodynamically favored one except at very low temperature.

We now focus our attention on the lattice thermal conductivity tensor of each structure. Under the relaxation-time approximation to the Boltzmann transport equation for phonons, it can be computed as

\begin{equation}
  \kappa^{\left(\mu\nu\right)} = \frac{1}{k_BT^2A}\sum\limits_\lambda n_{\lambda} \left(n_{\lambda} + 1\right)\left(h f_{\lambda}\right)^2 v_{\lambda}^{\left(\mu\right)}v_{\lambda}^{\left(\nu\right)} \tau_{\lambda}.
  \label{eqn:kappa}
\end{equation}

\noindent Here, $\mu$ and $\nu$ are Cartesian indices, $A$ is the area of the unit cell, $\boldsymbol{v}_{\lambda}$ is the group velocity of mode $\lambda$, and $\tau_{\lambda}$ its relaxation time. With this definition the S.I. units of $\boldsymbol{\kappa}$ are $\mathrm{W\,K^{-1}}$ instead of the more typical $\mathrm{W\,m^{-1}\,K^{-1}}$. This convention is better suited for 2D systems, where heat flux is defined per unit length, and not area. It enables us to dispense with introducing a thickness for the layers, and allows for easier comparison between different 2D systems.

It remains to compute the relaxation time in Eq. (\ref{eqn:kappa}). In the absence of crystalline defects, the dominant source of phonon scattering is anharmonicity, whose leading term can be characterized by the full set of third-order derivatives of the potential energy. This is particularly true for arsenic, a monoisotopic element, which lacks any contribution to phonon scattering from mass disorder \cite{tamura_isotope_1983}. We generate a minimal set of supercell configurations required to obtain all those constants, we obtain the forces on atoms in those configurations using VASP with the same parameters as for the harmonic calculations, and we rebuild the third-order force-constant tensor, compute the anharmonic scattering rates, and obtain $\boldsymbol{\kappa}$ from first principles using the almaBTE software package \cite{vermeersch_cross-plane_2016}, following the procedure described in detail in Ref. \onlinecite{cpc_2014}.

We also compute the reduced thermal conductivity \cite{carrete_nanograined_2014}:

\begin{equation}
  \tilde{\kappa}^{\left(\mu\nu\right)} = \frac{1}{k_BT^2A}\sum\limits_\lambda n_{\lambda} \left(n_{\lambda} + 1\right)\left(h f_{\lambda}\right)^2 \frac{v_{\lambda}^{\left(\mu\right)}v_{\lambda}^{\left(\nu\right)}}{\left\vert \boldsymbol{v}_{\lambda}\right\vert}.
  \label{eqn:reduced}
\end{equation}

\noindent If every heat carrier in the system had the same mean free path $\Lambda = \left\vert\boldsymbol{v}_{\lambda}\right\vert\tau_{\lambda}$, $\boldsymbol{\kappa}$ would become proportional to $\Lambda$, with $\tilde{\boldsymbol{\kappa}}$ as the proportionality constant. Hence, $\tilde{\boldsymbol{\kappa}}$ can be interpreted as a purely harmonic descriptor of the effect of phonon frequencies and group velocities on the thermal conductivity when anharmonicity is homogenized in a particular way.

Based on symmetry considerations, the buckled and s/o arsenene monolayers must be isotropic, with a scalar thermal conductivity. The puckered layer, in contrast, has two independent thermal conductivities corresponding to the nonequivalent zigzag and armchair directions.

\begin{figure}
  \begin{center}
    \includegraphics[width=\linewidth]{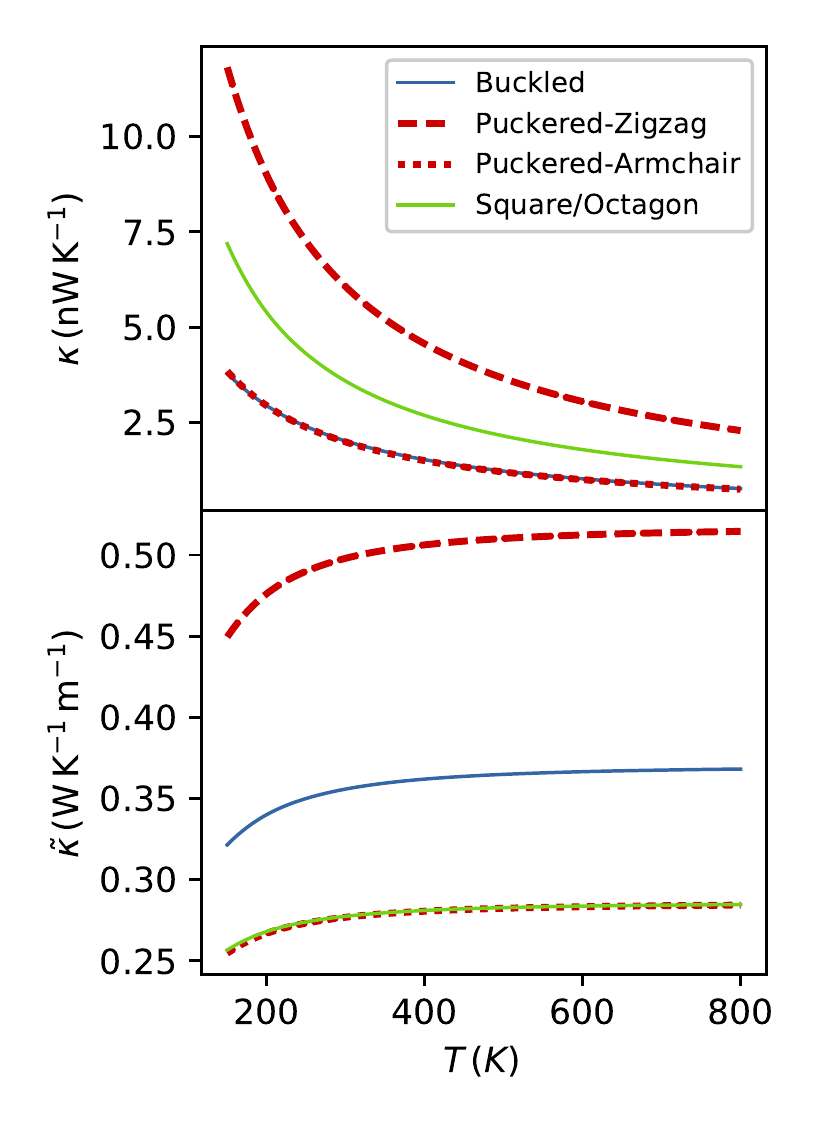}
  \end{center}
  \caption{Thermal conductivity (top) and reduced thermal conductivity (bottom) for the three stable phases of arsenene studied in this work, as a function of temperature.}
  \label{fig:kappa}
\end{figure}

The independent components of $\boldsymbol{\kappa}$ and $\tilde{\boldsymbol{\kappa}}$ are shown in Fig. \ref{fig:kappa}. The puckered phase is confirmed as highly anisotropic\cite{zeraati_highly_2016}. More remarkably, those results show that the buckled phase conducts heat more poorly than any of the others. This observation contradicts the intuition that to maximize lattice thermal conductivity one should look for simple structures, with high band degeneracy and few possible phonon scattering routes. Here, even the $8$-atom-per-unit-cell s/o structure displays roughly twice the room-temperature thermal conductivity of the simpler buckled phase.

A good understanding of the factors determining whether a material is a good or a bad thermal conductor is critical when looking for innovative solutions to heat dissipation problems. Hence we take a deeper look at the reasons for the counter-intuitive ordering of the thermal conductivities of arsenenes. The bottom panel in Fig. \ref{fig:kappa} gives a hint that the causes behind the high thermal conductivity of the two more complex phases are different. Specifically, the reduced thermal conductivity of puckered arsenene in the zigzag direction is already much higher than that of the buckled phase, but the value for the s/o structure is actually much lower. In other words, harmonic effects are responsible for the high conductivity of the puckered structure. In the s/o structure, however, a reduced anharmonicity plays the key role in preventing $\boldsymbol{\kappa}$ from falling much lower. 

Regarding the puckered phase, the favorable harmonic contribution to thermal transport may be due to the low-lying optical branches with relatively high average velocities in the direction of reciprocal space corresponding to the short axis of the puckered-arsenene unit cell (Fig. \ref{fig:phonons}, top-right panels). As for the s/o phase, a reduced thermal conductivity is to be expected given the generally low group velocities of low-frequency phonons (Fig. \ref{fig:phonons}, bottom-right panels), while the low scattering rates may be due to a combination of the high symmetry of the phase plus the very localized character of the two groups of high-frequency modes, which could have acted as scattering channels had they had a greater overlap with those in the low-energy region.

To sum up, we characterize the ground state and phonon dispersions of four proposed arsenene structures (buckled, puckered, tricycle and square/octagon) with very different geometric features and degrees of structural complexity. Our analysis shows the tricycle structure to be unstable. Furthermore, the relative stability of the remaining three can only be correctly understood when phonons are taken into account, even at vanishing temperature. Specifically, we predict the puckered and s/o structures to be thermodynamically favored at low and high temperatures respectively, in contrast with previous assertions in the literature that the buckled phase would be the most stable. We then characterize also the anharmonic features of the three stable structures and compute their thermal conductivities from first principles. We show the two most complex structures to have high thermal conductivities due to a combination of harmonic and anharmonic features, which eludes simplistic interpretations in terms of unit cell complexity.

We expect our results to be of help to experimental efforts toward the synthesis of proposed arsenene structures. More generally, every conclusion from this study highlights the crucial role of an appropriate treatment of phonons in the equilibrium and transport thermodynamics of 2D systems. This can be expected to become more important as computational explorations of libraries of 2D materials become more elaborate, since more numerous, more complex and chemically richer structures can be expected to be put forward, demanding careful examination of their practical feasibility.

\begin{acknowledgement}
The authors acknowledge the support from the European Union's Horizon 2020 Research and Innovation Programme [grant no. 645776 (ALMA)] and from the Xunta de Galicia (grants nos. AGRUP2015/11 and GRC ED431C 2016/001) in conjunction with the European Regional Development Fund (FEDER). We also thank Dr. Amador Garc\'ia-Fuente for useful discussions. The computational results presented have been achieved in part using the Vienna Scientific Cluster (VSC).
\end{acknowledgement}

\begin{suppinfo}
  Contributions to the thermal conductivity from each branch for the three stable phases of arsenene, along with phonon scattering rates, are given in the Supporting Information.
\end{suppinfo}

\bibliography{arsenene}

\end{document}